\documentclass[twocolumn,aps,prl]{revtex4}
\usepackage[english]{babel}
\usepackage[dvipdf]{graphicx}
\usepackage{amssymb}
\usepackage{amsmath}
\usepackage{amsbsy}
\usepackage{dcolumn}

\usepackage{color}

\begin{document}

\title{Visco-elasticity of colloidal polycrystals doped with impurities}

\author{Ameur Louhichi, Elisa Tamborini, Julian Oberdisse, Luca Cipelletti and Laurence Ramos$^{\star}$ }

\affiliation{
Laboratoire Charles Coulomb (L2C), UMR 5221 CNRS-Universit\'e de Montpellier, Montpellier, F-France}

\email{laurence.ramos@umontpellier.fr}
\date{\today}

\begin{abstract}

We investigate how the microstructure of a colloidal polycrystal influences its linear visco-elasticity. We use thermosensitive copolymer micelles that arrange in water in a cubic crystalline lattice, yielding a colloidal polycrystal. The polycrystal is doped with a small amount of nanoparticles, of size comparable to that of the micelles, which behave as impurities and thus partially segregate in the grain boundaries. We show that the shear elastic modulus only depends on the packing of the micelles and does not vary neither with the presence of nanoparticles nor with the crystal microstructure. By contrast, we find that the loss modulus is strongly affected by the presence of nanoparticles. A comparison between rheology data and small-angle neutron scattering data suggests that the loss modulus is dictated by the total amount of nanoparticles in the grain boundaries, which in turn depends on the sample microstructure.
\end{abstract}
\maketitle

\section{Introduction}

Most solid materials in everyday life, as e.g. metals or ceramics, are crystals, whose constituents (atoms or molecules) are regularly positioned in space. Over the past century scientists have reached a good understanding of the electric, thermal, and mechanical properties of a defect-free crystal. However, real-life crystalline materials have defects, which strongly affect their properties~\cite{Howe1997}. In particular, crystalline solids such as metals and ceramics are usually aggregates of crystalline grains that form the \textit{microstructure} of the material, which plays a key role in almost all industrial processes and products.

Very generally, the mechanical properties of crystalline materials strongly depend on defects in the crystalline structure. The interplay between defective structure and mechanical properties has been extensively investigated in metals, and more recently in nanocrystalline materials, i.e. materials where the size of the grains is smaller than typically $100$ nm, because of their outstanding chemical and mechanical properties, including superplasticity~\cite{Lu2000, Schiotz1998, Schiotz2003, Yamakov2004, Pan2007, Li2008, Wang2010, Chookajorn2012}. Grain boundaries (GBs), the two-dimensional lattice defects that separate the different grains, control the bulk mechanical properties of polycrystalline materials~\cite{Lu2000, Schiotz1998, Schiotz2003, Yamakov2004, Li2008, Shan2004, Buban2006, Cheng2009}.
In particular, the sliding and the migration of GBs play important roles in the plastic deformation and the fracture at high temperature.

How GBs and defects in general impact the mechanical properties of soft materials is far less documented. A few exceptions include liquid crystalline materials where inclusion of particles has been shown to induce defects that strongly modify the linear mechanical properties of the composite due to the presence of a stabilized network of defects~\cite{Zapotocky1999,Wood2011}. The link between microstructure and mechanical properties in colloidal materials was invoked in a few previous studies with dispersions of block copolymers \cite{Eiser2000,Eiser2000b,Pozzo2005,Torija2011} but was not investigated in detail and we are not aware of any experimental work that investigates how the visco-elasticity of a colloidal polycrystal depends on its microstructure, i.e. on the average size of the crystalline grains. A prerequisite for such study is to be able to tune the average size of the crystalline grains, which in atomic and molecular crystals is typically achieved by varying the cooling rate. For colloidal crystals, this is not an easy task~\cite{Palberg1995, Engelbrecht2010, Gokhale2013} because the crystallization rate is controlled by the volume fraction of colloids, a parameter that usually cannot be varied during an experiment, unlike temperature $T$ for hard condensed matter systems.

This difficulty is overcome in a novel colloidal system that we have recently developed and that displays analogies with atomic and molecular crystals. First, thanks to the thermosensitivity of the colloids, the crystallization can be induced at a controlled rate by changing temperature \cite{Peng2010,Ghofraniha2012,Gokhale2012}. Second, the system comprises small amounts of nanoparticles of physico-chemical nature different from that of the colloids forming the crystal, which act as impurities in atomic and molecular systems \cite{Lejcek2010, Losert1998} and as such partially segregate in the grain boundaries during the crystallization process \cite{Villeneuve2005, Yoshizawa2011, Ghofraniha2012}. The microstructure can be controlled both by the rate at which the temperature is varied, and by the amount of nanoparticles.  We leverage on these specificities to investigate by rheology how the microstructure influences the sample visco-elasticity. Complementary neutron scattering data are used to understand the microscopic origin of the observed mechanical behavior.

The paper is organized as follows. We first describe the materials and experimental methods. We then present how the linear visco-elasticity varies with the amount of nanoparticles and the temperature rate used to prepare the samples. Complementary experiments by scattering techniques are shown in order to relate our findings on the visco-elasticity to the sample microstructure. We finally discuss our results in light of the knowledge on atomic and molecular crystals, and on other soft materials with a cellular structure.

\section {Materials and Methods}
\subsection{Experimental samples}

The colloidal crystals have been described elsewhere~\cite{Tamborini2012,Louhichi2013,Tamborini2014}. They comprise a commercial block-copolymer, water and nanoparticles.  We use a thermosensitive triblock copolymer of Pluronics type (F108), made of a central polypropylene oxide (PPO) block of $52$ monomers flanked at each extremity by two polyethylene oxide (PEO) blocks of $132$ monomers each. The concentration of F108 in water is set at $34$\% w/w. The affinity of the PPO block for water varies with temperature. At low temperature ($T \lesssim 6^{\circ}\mathrm{C}$), the copolymer is fully water-soluble and the sample is a viscous fluid (zero-shear viscosity $0.32$ Pa.s at $T = 4^{\circ}\mathrm{C}$). Conversely, at room temperature the affinity of PPO for water is poor, and F108 self-assembles into micelles of diameter $2a=22$ nm that arrange on a face-centered cubic lattice (lattice parameter $32$ nm)~\cite{Alexandridis1995, Tamborini2012}. A small amount of nanoparticles (NPs) is added to the F108/water mixture. We used Bindzil silica NPs (kind gift from Eka Chemical, sample type 40/130), with an average diameter of $30$ nm and a relative polydispersity of $19\%$, as determined by transmission electron microcopy. A few complementary experiments are performed with carboxylated polystyrene nanoparticles of diameter $36$ nm purchased from Invitrogen. The volume fraction of NPs, $\phi$, is varied in the range $(0.5-2)\%$. The sample is prepared in a fluid phase at $T \approx 3^{\circ}\mathrm{C}$. Crystallization is then induced by rising $T$ at a fixed heating rate, with $\dot{T}$ in the range $(0.001-2)^{\circ}$C/min. We have previously shown that the crystallization temperature continuously increases with the heating rate from $15.5$ to $18.5^{\circ}$C~\cite{Louhichi2013}. Both rheology and scattering experiments are performed at $T=23^{\circ}\mathrm{C}$, where all samples are fully crystallized.

For neutron scattering measurements, water ($H_2O$) is replaced by an equal volume of deuterated water ($D_2O$) or a mixture of $D_2O$ and $H_2O$, with various ratios, in order to match the scattering of the solvent with that of the copolymer (resp., with that of the NPs), so that only the NPs (resp., only the micelles) should contribute to the scattering signal. Samples without NPs are prepared with pure $D_2O$ in order to maximize the contrast between the copolymer and the solvent. Samples with NPs are prepared using $39/61$ $H_2O/D_2O$ w/w to contrast-match the NPs or $85/15$ $H_2O/D_2O$ w/w to contrast-match the copolymer, as detailed previously~\cite{Tamborini2012}. We have checked that changing the solvents does not modify the sample structure~\cite{Tamborini2012}.

\begin{figure}
\includegraphics[width=0.45\textwidth]{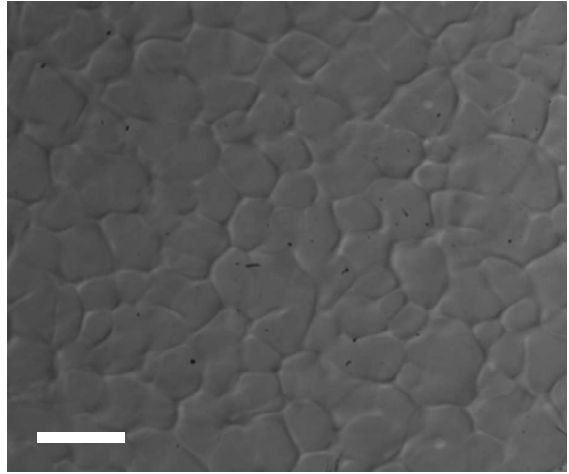}
\caption{Differential interference contrast microscopy image of a sample comprising nanoparticles with a volume fraction $\phi=2$\%, and prepared with a heating ramp $\dot{T}=0.001 \, ^{\circ}$C/min. Scale: $20 \, \mu$m.}
\label{fig:microscopy}
\end{figure}

The NPs act as impurities in the colloidal crystal. As such, they are found to partially segregate in the grain boundaries upon crystallization. As an illustration, we show in fig.~\ref{fig:microscopy} a light microscopy image of a sample comprising nanoparticles. Thanks to the accumulation of NPs in the GBs, the contrast between the crystalline grains and the GBs is enhanced and allows for the imaging of the network of GBs that delineate grains with different crystalline orientations. For the sample shown in fig.~\ref{fig:microscopy} ($\dot{T}=0.001^{\circ}$C/min, $\phi=2$ \%), the average size of the crystalline grains is of the order of $10 \, \mu$m, but the sample microstructure can be tuned by varying the amount $\phi$ of NPs, or the heating rate $\dot{T}$~\cite{Ghofraniha2012}. The average size of the crystallites typically ranges between $8$ and $45 \, \mu$m.

Using fluorescent polystyrene nanoparticles of size comparable to that of the silica NPs used here, and quantitative confocal imaging, we have previously shown~\cite{Ghofraniha2012} that not all NPs are located in the GBs. A partitioning coefficient, $\mathcal{P}$, can be introduced, defined as the ratio between the total amount of NPs in the GBs and the total amount of NPs inside the grains. We have found~\cite{Ghofraniha2012} that $\mathcal{P}$ is not constant but rather decreases as the crystallite size becomes smaller. Interestingly, $\mathcal{P}$ only depends on the microstructure and not on what combination of $\phi$ and $\dot{T}$ is used to achieve a given crystallite size.

\subsection{Experimental techniques}

The linear visco-elasticity of the samples is investigated using a stress-controlled rheometer (USD 200, Physica) operating in a strain-controlled mode thanks to a feedback loop. The sample, initially in a liquid phase at $T=3^{\circ}$C, is loaded in a Couette cell whose temperature is also set at $3^{\circ}$C, and is covered with a thin layer of silicon oil to prevent evaporation. The sample temperature is increased at a fixed temperature rate $\dot{T}$ using a water circulating thermostat bath. Visco-elasticity measurements are performed by imposing an oscillatory strain of amplitude $\gamma$ and frequency $f$ and by measuring the complex shear modulus, $G^* = G' + iG''$~\cite{Larson1999}.

Additional experiments are performed using small-angle neutron scattering. Details of the experimental conditions are described elsewhere~\cite{Tamborini2012}. In brief, experiments are run on the beamline PACE at the Laboratoire Leon Brillouin, Saclay, France. To cover the range of scattering vectors $q\in [(0.003-0.4)\AA^{-1}]$, three experimental configurations are used, $D=4.5$ m, $\lambda=12 \, {\AA}$ ; $D=4.5$ m, $\lambda=6 \, {\AA}$, $D=1$ m, $\lambda=6 \, {\AA}$, where $D$ is the sample-detector distance and  $\lambda$ is the neutron wavelength. Here, $q=4\pi\lambda^{-1}\sin(\theta/2)$  is the magnitude of the scattering vector and  $\theta/2$ is the scattering angle.
Empty cell background subtraction, calibration by light water in a $1$ mm-thick quartz cell, and absolute determination of scattering intensities $I(q)$  per sample volume are performed using standard procedures~\cite{Lindner2002,Tamborini2012}.

\section{Results and discussion}

\subsection{Visco-elasticity of samples without nanoparticles}

\begin{figure}
\includegraphics[width=0.45\textwidth]{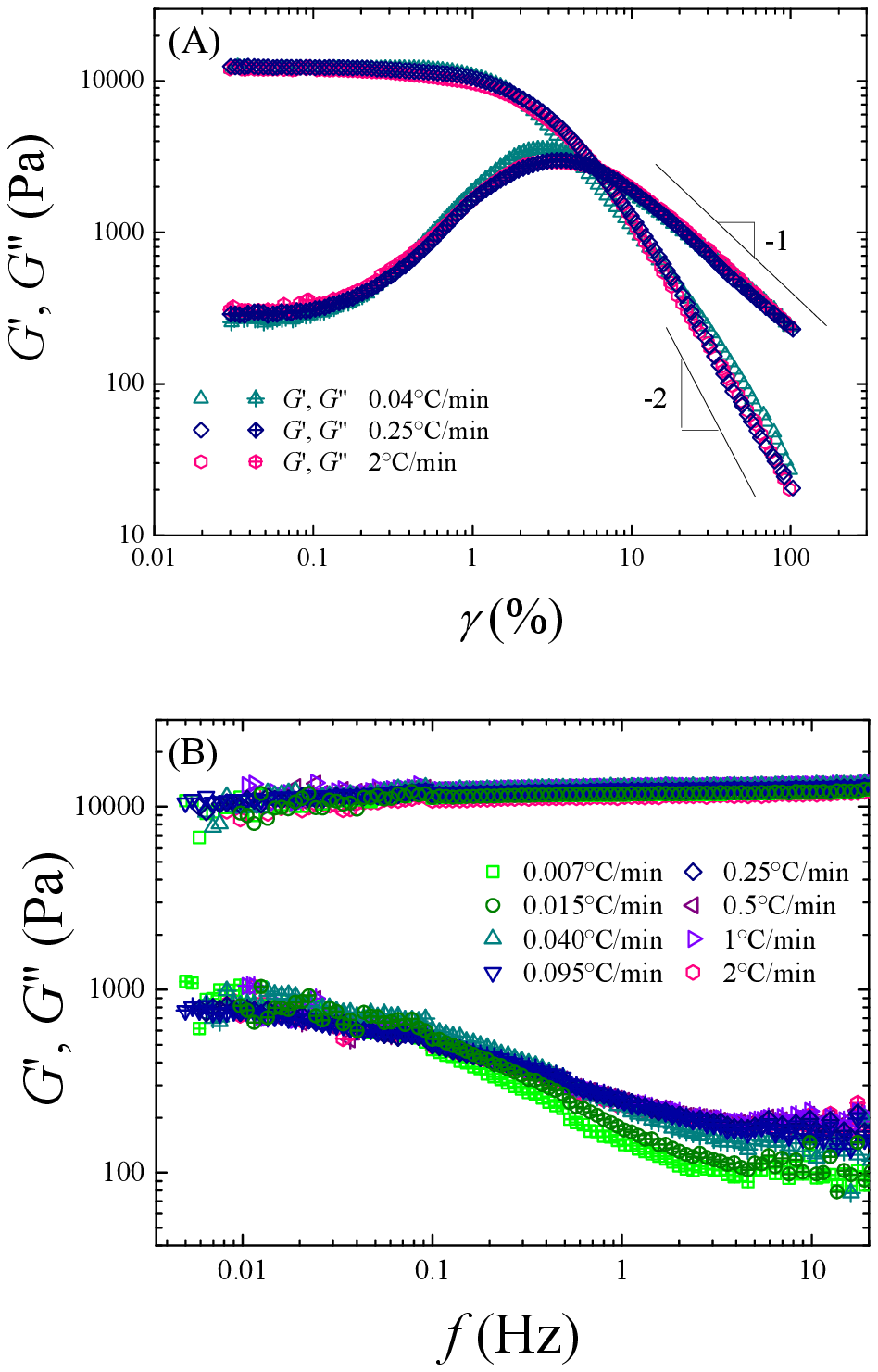}
\caption{(Color online). Strain sweeps (A) and frequency sweeps (B) for samples without nanoparticles, prepared using various different temperature ramps as indicated in the legend. The frequency is $0.5$ Hz in the strain sweep experiments and the strain amplitude is $0.1 \%$ in the frequency sweep experiments.}
\label{fig:WoNP}
\end{figure}

We first investigate the linear visco-elasticity of samples without nanoparticles. Strain sweeps measured at a frequency of $f=0.5$ Hz and frequency sweeps measured in the linear regime ($ \gamma=0.1 \% $) are shown in fig.~\ref{fig:WoNP} for samples prepared with different heating rates $\dot{T}$. Overall, the visco-elasticity is only very weakly dependent on $\dot{T}$. The $\gamma$ dependence of the complex modulus exhibits the hallmarks of the visco-elasticity of soft solids, as found for many materials ranging from colloidal gels to concentrated emulsions and dry foams~\cite{Wyss2007}.  In the linear regime, below $\gamma \simeq 0.2$\%, the storage modulus $G'$ and the loss modulus $G"$ are independent of the applied strain. Above a critical strain of order 10\%, $G' \propto \gamma^{-2}$. The loss modulus, by contrast, exhibits a well defined peak, whose maximum occurs roughly at the cross-over of $G'$ and $G"$, before decreasing proportionally to $\gamma^{-1}$. The frequency-dependent complex modulus in the linear regime is dominated by the storage modulus, which, in the range of frequency investigated [($0.005-20$) Hz], is between one and two orders of magnitude larger than the loss modulus, indicating a highly elastic soft material. We find that the storage modulus is nearly frequency-independent over the whole range of probed frequencies. Moreover, the data for $G'$ acquired at different heating rates $\dot{T} \in (0.007-2)^{\circ}$C/min perfectly superimpose, yielding a frequency- and heating rate-independent elastic plateau, $G_0= (11610 \pm 640)$ Pa (fig.~\ref{fig:Comparison}A). We find on the other hand that the loss modulus continuously decreases from about $1000$ Pa down to about $100$ Pa as the frequency increases from $0.005$ to $20$ Hz. At relatively high frequency ($f>0.5$ Hz) the loss modulus slightly varies with the heating ramp. At $f=2$ Hz, $G"$ increases from $110$ to $215$ Pa and finally reaches a plateau for $\dot{T}\ge 0.1^{\circ}$C/min (circles in fig.~\ref{fig:Comparison}B).

\subsection{Visco-elasticity of samples doped with nanoparticles}

\begin{figure}
\includegraphics[width=0.45\textwidth]{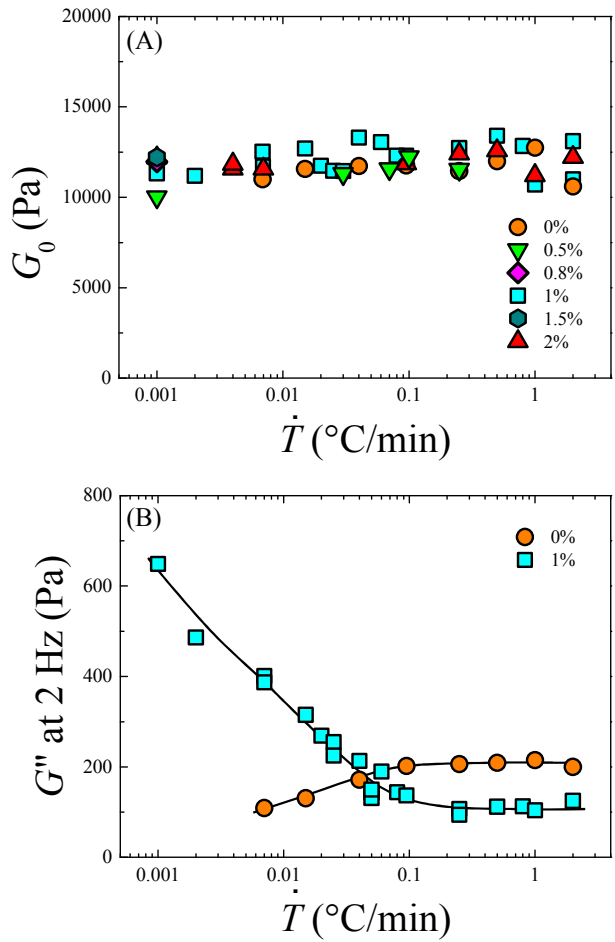}
\caption{(Color online). Elastic plateau (A) and loss modulus (B), as a function of the temperature ramps used to prepare the samples. Samples without nanoparticles and with various amounts of nanoparticles, as indicated in the caption, are compared. The loss modulus has been measured at $2$ Hz. The lines are guides for the eye.}
\label{fig:Comparison}
\end{figure}

\begin{figure}
\includegraphics[width=0.45\textwidth]{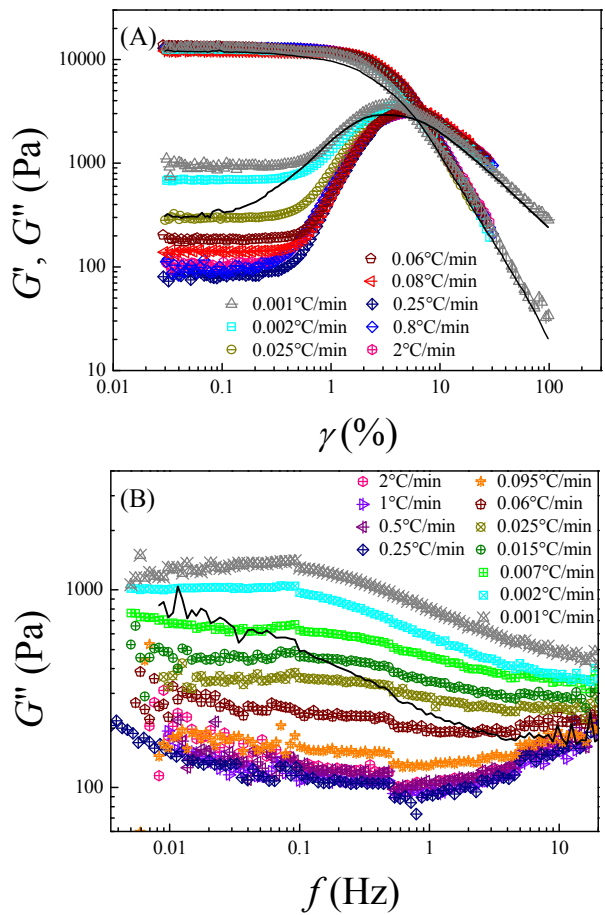}
\caption{(Color online). Strain sweeps (A) and frequency sweeps (B) for samples with $1 \, \%$ nanoparticles, prepared using various temperature ramps, whose rate $\dot{T}$ is indicated in the legend (symbols). $f=0.5$ Hz in the strain sweep experiments and $\gamma = 0.1 \%$ in the frequency sweep experiments. In (B) only the loss modulus is plotted. The continuous lines correspond to the data for a sample without nanoparticles, and with $\dot{T} = 2~^{\circ}\mathrm{C}$/min.}
\label{fig:NP}
\end{figure}

The elastic modulus of samples doped with nanoparticles behaves closely to that of samples without NPs, in that $G'$ is both frequency-independent (data not shown) and nearly independent of $\dot{T}$ (fig.~\ref{fig:Comparison}A). Moreover, the numerical value of the elastic modulus is insensitive to the content of NPs: in the explored range $0 \le \phi \le 2\%$, all $G_0$ vs $\dot{T}$ data collapse almost perfectly on the same master curve as for samples without NPs, as seen in fig.~\ref{fig:Comparison}A. Finally, the way the storage modulus depends on $\gamma$ in the non-linear regime is similar to that for a sample without NPs. This is illustrated by the superimposition of the data with and without NPs for $\gamma \geq 3$ \%, i.e. in the regime where $G" > G'$ (fig.~\ref{fig:NP}A). As seen in fig.~\ref{fig:NP}A, this also holds for the loss modulus $G''$. Hence, the addition of NPs does not alter the elastic properties of our polycrystals in both the linear and non-linear regimes, nor does it modify the viscous properties in the large-amplitude, non-linear regime.

By contrast, in the linear regime the loss modulus exhibits significant variations between samples with and without NPs. We first discuss the differences between a pure polycrystal and samples doped with a fixed amount of NPs and prepared at various ramp rates $\dot{T}$. Note that we describe the data obtained with $\phi = 1\%$, but the main conclusions are robust with respect to the amount of added NPs. First, in sharp contrast to the behavior of samples without NPs, the strain sweeps performed at $ f=0.5$ Hz show that, at low $\gamma$, $G"$ strongly increases as the heating rate decreases (fig.~\ref{fig:NP}A).
We plot in fig.~\ref{fig:Comparison}B $G''(f=2~\textrm{Hz})$ as a function of the temperature ramp rate. The loss modulus decreases when the heating rate increases and reaches a plateau for $\dot{T} \gtrsim 0.1^{\circ}\mathrm{C}$. This trend is robust with respect to the choice of the frequency at which the loss modulus is measured: as it can be inferred from fig.~\ref{fig:NP}B, a decrease of $G''$ followed by a plateau at $\dot{T} \gtrsim 0.1^{\circ}\mathrm{C}$ is systematically seen in the whole range of accessible frequencies. (An example for data acquired at $f=0.05$ Hz will be shown in fig.~\ref{fig:RheoPlusSANS}A below). Within this general trend, only the amplitude of the decrease of $G''$ with $\dot{T}$ changes, the largest drop of the loss modulus being observed at low frequency (e.g. compare the drop by a factor of $11$ at $f=0.05$ Hz, to the decrease by a factor of $6$ at $f=2$ Hz, see fig.~\ref{fig:NP}B). Another remarkable feature is that the way the linear loss modulus varies with frequency in samples with NPs differs from that without nanoparticles. In both cases, the loss modulus decreases when the frequency increases, except at large frequency ($f$ larger than ~$1$ Hz), where $G"$ eventually reaches a plateau or slightly increases. However, we find that the evolution of $G"$ with frequency is much smoother in the presence of NPs (fig.~\ref{fig:NP}B).

Overall, these data show that the behavior of the loss modulus is rather complex. Naively, one would expect that dissipation increases as the crystal contains more defects. When the heating rate increases, the crystallization process is faster, yielding smaller crystallites and more defects. Thus, one would expect larger $G"$ values as $\dot{T}$ increases. For samples without NPs, we indeed measure a continuous increase of $G"$ with $\dot{T}$ (circles in fig.~\ref{fig:Comparison}B). Although the effect is rather weak and only detectable at high frequency, the measured trend is in agreement with the physical expectation. The results for samples with NPs are in sharp contrast with this picture. We find that dissipation is strongly influenced by the way the samples are prepared. Surprisingly, we observe a marked decrease of $G"$ when $\dot{T}$ increases (squares in fig.~\ref{fig:Comparison}B), a trend opposite to that measured without NPs. Microscopy and scattering experiments~\cite{Tamborini2012,Louhichi2013} indicate that when $\dot{T}$ increases the average size of the crystallite decreases. Hence, the total surface of grain boundaries, where we expect dissipation to occur predominantly, increases. We are then left with the puzzling observation of a growth of the regions where dissipative processes tend to concentrate, concomitant with a decrease of the loss modulus.

\subsection{Correlation with structural data}

\begin{figure}
\includegraphics[width=0.45\textwidth]{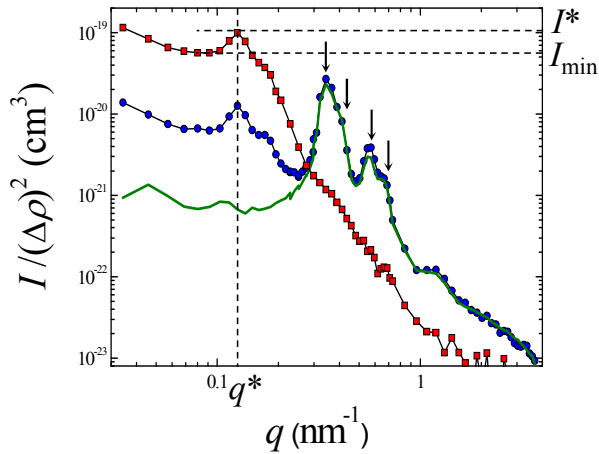}
\caption{(Color online). Small-angle neutron scattering intensity profiles for samples prepared with a temperature ramp of $\dot{T}=0.007 \, ^{\circ}$C/min. The intensity is normalized by the scattering contrast $\Delta \rho$ between the solvent and the scattering objects (polymer, blue circles, and NPs, red squares). The green continuous line is a sample without nanoparticles (NPs) and the symbols correspond to samples comprising $1 \, \%$ NPs under different scattering contrast conditions. Red squares: polymer-matched sample, only the micelles contribute to $I(q)$. Blue circles: silica-matched sample. The dotted
lines indicate the position of the structure peak of the NPs, at $q^*$, due to their confinement in the grain-boundaries, and the intensity $I^*$ and $I_{\min}$ used to measure the peak amplitude. The arrows indicate the peaks due to the scattering of the micellar crystal.}
\label{fig:SANS}
\end{figure}

\begin{figure}
\includegraphics[width=0.45\textwidth]{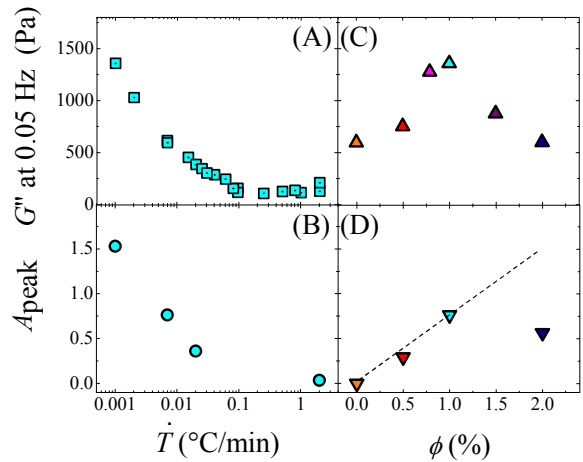}
\caption{(Color online). Comparison between the linear visco-elasticity measurement and the structural measurements. In (A,C), the loss modulus taken at $2$ Hz is plotted as a function of the heating ramp used to prepared the samples, $\dot{T}$ (A), and as a function of the concentration of nanoparticles, $\phi$ (C). In (B,D), the relative amplitude of the NPs structure peak is plotted as a function of $\dot{T}$ (B), and as a function of $\phi$ (D). In (A,B) $\phi=1 \, \%$. In (C), $\dot{T}=0.001 \, ^{\circ}$C/min, and in (D), $\dot{T}=0.02 \, ^{\circ}$C/min. In (D), the dotted line is a guide for the eye.}
\label{fig:RheoPlusSANS}
\end{figure}

These counterintuitive results can be rationalized by considering the structure of the samples at length scales comparable to the size of the nanoparticles. Figure~\ref{fig:SANS} shows the scattering intensity profiles of several samples with and without NPs, comprising different mixtures of $H_2O$ and $D_2O$, but all prepared with a same heating rate ($\dot{T}=0.007 \, ^{\circ}$C/min). The sample without NPs exhibits structural peaks due to the crystalline order of the micelles in the material (green line and arrows in fig.~\ref{fig:SANS}). The same peaks are also found for samples with NPs, thus demonstrating that NPs do not modify the crystalline packing (blue circles).  In the presence of NPs, an additional peak located at $q^*=0.126 \, \rm{nm}^{-1}$ is detected. In experimental conditions where the scattering length density of the polymer is matched to that of the solvent, the signal of the colloidal crystal vanishes, and only the signal from the NPs is visible. In this case, the peak at $q^*$ is still present, demonstrating that it originates from the structure of the NPs in the colloidal crystal. The peak position yields an average distance between the NPs, $2\pi/q^* = 50$ nm. This length scale is slightly larger than the NP diameter ($30$ nm) and corresponds to an effective volume fraction of the NPs of the order of $(10-15)$\%, hence a concentration very large compared to the average concentration of NPs in the sample ($1$\%). We interpret this peak as the signature of the packing of the NPs in the grain boundaries. The position of this peak does not vary significantly with $\phi$ or $\dot{T}$, indicating a nearly constant average distance between NPs in the GBs~\cite{Tamborini2012}. By contrast, its amplitude changes. Note that the amplitude of the peak is only dictated by the total number of NPs in the GBs, as the NPs in the bulk of the crystalline grains are too diluted  to contribute to the scattering intensity in the range of wave vectors investigated here. We quantify the relative amplitude of the peak by $A_{\rm{peak}}=(I^* - I_{\rm{min}})/I_{\rm{min}}$, where $I^*$ is the scattered intensity measured at the peak position, $q^*$, and $I_{\rm{min}}$ is the minimum
scattered intensity measured at $q$ lower than $q^*$, and plot the evolution of $A_{\rm{peak}}$ with $\dot{T}$ and $\phi$ in figs.~\ref{fig:RheoPlusSANS}B and D. For samples comprising a fixed amount of nanoparticles, we find that $A_{\rm{peak}}$ decreases as $\dot{T}$ increases (fig.~\ref{fig:RheoPlusSANS}B). Our data thus indicate that for these samples the total amount of NPs in the GBs decreases when $\dot{T}$ increases. However, as $\dot{T}$ increases the average crystallite size decreases, and thus the total surface area of GBs increases. The fact that the cumulated amount of NPs in the GBs decreases, in spite of an increase of the total surface area of the grain boundaries, is due to a change in the efficiency of the segregation process of the NPs during crystallization. Indeed, as already briefly mentioned in the Materials and Methods section and as described in detail elsewhere~\cite{Ghofraniha2012}, not all the NPs incorporated in the polycrystal are located in the grain boundaries: the partitioning coefficient $\mathcal{P}$ depends on the sample microstructure. For fluorescent polystyrene NPs of size comparable to the silica NPs used here, we have previously found that $\mathcal{P}$ diminishes as the average size of the crystallites decreases. Although the NPs investigated here are of different chemical nature, it is likely that a similar behavior also holds for the silica NPs. We indeed found that the evolution of the grain size with $\phi$ and $\dot{T}$ are similar for both kinds of particles, although the grains are slightly larger for silica NPs than for polystyrene NPs. Additionally, a similar dependence on crystallization rates of the partitioning coefficient is commonly observed in atomic and molecular systems~\cite{Aziz1986}.
The data shown in fig~\ref{fig:RheoPlusSANS}B suggest that, in the linear regime, dissipation is mostly due to the NPs that are segregated in the grain boundaries. Within this scenario, the decrease of $G"$ with $\dot{T}$ would be the consequence of reduced dissipation stemming from a smaller number of NPs in the grain boundaries as $\dot{T}$ increases, due to the reduction of the partitioning coefficient.

To further test these ideas, we perform rheology and SANS measurements on a series of samples with various amounts of NPs, keeping fixed the heating rate. Data obtained by SANS (fig.~\ref{fig:RheoPlusSANS}D) show a non-monotonic behavior of $A_{\rm{peak}}$ with $\phi$. This can be interpreted as the result of the competition between two opposite effects: on the one hand, the higher $\phi$, the larger the number of NPs that may be segregated in the GBs. On the other hand, the higher $\phi$ the lower the partition coefficient and hence the lower the fraction of NPs that actually accumulate in the GBs rather than in the bulk of the crystallites. At small $\phi$, the first effect dominates: the size of the grains and the partition coefficient $\mathcal{P}$ do not vary significantly with $\phi$, implying that the cumulated number of NPs in the GBs is directly proportional to the concentration of NPs. Consequently, $A_{\rm{peak}} \sim \phi$, consistently with the experimental data (dotted line in fig.~\ref{fig:RheoPlusSANS}D). As $\phi$ increases further, the average size of the crystallite decreases and the partitioning becomes less efficient, leading to a decrease of the relative amount of NPs in the grain-boundaries. This explains why $A_{\rm{peak}}$ stops growing and eventually slightly decreases at higher $\phi$. Remarkably, a similar non-monotonic behavior with $\phi$ is observed also for the loss modulus (fig.~\ref{fig:RheoPlusSANS}C), leading support to the hypothesis that $G"$ is dominated by the dissipation associated with the nanoparticles in the grain-boundaries.

\subsection{Discussion}

We have found that the linear storage modulus is nearly frequency independent and constant irrespective of the presence or not of nanoparticles and of the heating rate used to prepare the samples, within experimental uncertainties. An average over all measurements yields $G_0=(11970\pm750)$ Pa. Consistently with previous results obtained in comparable conditions \cite{Pozzo2005}, we do not measure any weakening of the elasticity due to the presence of NPs. The elasticity is independent of the presence of NPs and of the heating rate, and thus on the crystal microstructure. Hence, the elasticity is governed solely by the packing of the micelles in the crystalline lattice. Using simple scaling arguments,  $G_0$ is expected to scale as $k_B T / a^3$, where $a=11$ nm is
the radius of the micelles. We find $G_0 a^3 /k_B T=3.9$, a value in good agreement with experiments on
colloidal hard sphere crystals, for which, e.g., $G_0 a^3 /k_B T=5.8$ at a colloid volume fraction of $0.55$~\cite{Phan1999}.

The behavior of the loss modulus is much richer. On the one hand, the physical origin of the dissipation with and without nanoparticles is different, as reflected by the very different dependence of $G"$ with frequency in the two cases. For samples without NPs, the large values of the loss modulus at small frequency indicate dissipation processes with long characteristic relaxation times, presumably due to point and line defects in the bulk of the crystallites, since $G''$ is insensitive to the amount of GBs at low frequency. Samples with NPs and prepared at a high heating rate certainly contain a large number of bulk defects, since the NP impurities tend to be dispersed throughout the sample, rather than massively accumulate in the GBs.  One might then expect that $G''$ would be as high or even higher than in the NPs-free samples. Surprisingly, in the presence of nanoparticles the opposite trend is observed: $G''$  has lower values and is almost frequency-independent. We propose that this stems from a stabilizing effects of the NPs, leading to a slowing down of the defect dynamics, as observed in colloidal crystals~\cite{Villeneuve2005, Villeneuve2009} and in other materials ~\cite{Hazzledine1990, Soer2004, Li2014}.

\begin{figure}
\includegraphics[width=0.45\textwidth]{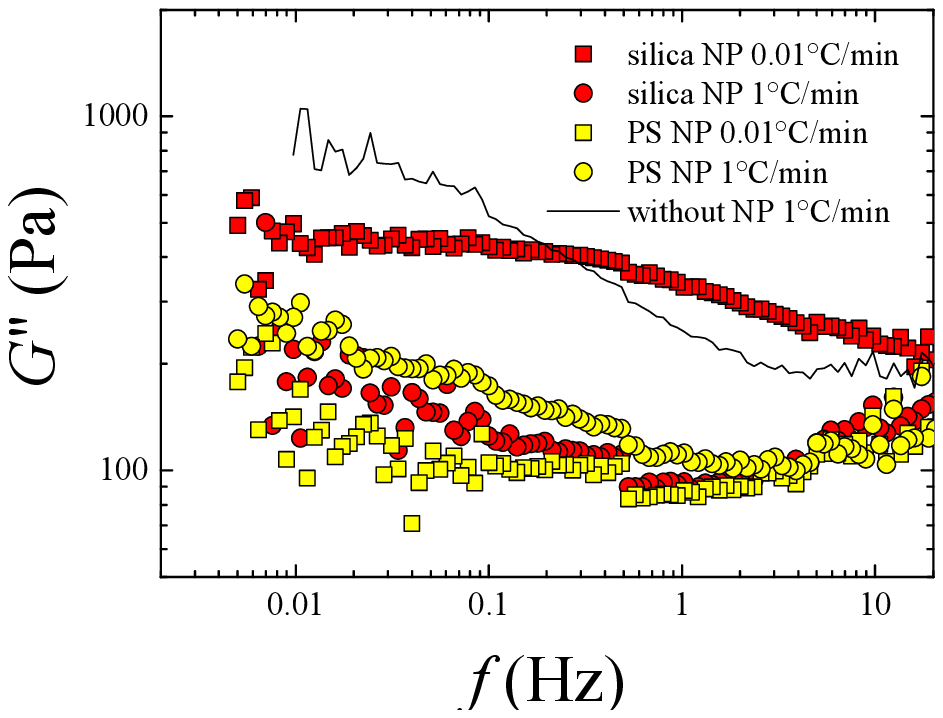}
\caption{(Color online). Frequency dependence of the loss modulus for samples with nanoparticles, prepared using two different temperature ramps, $1 \, ^{\circ}$C/min (circles) and $0.001 \, ^{\circ}$C/min (squares). Two kinds of particles are used: silica NPs (red, dark gray, symbols) and polystyrene (PS) (yellow, light gray, symbols). The volume fraction of NPs is $\phi=0.5$ \%. The strain amplitude is $0.1 \%$. For the sake of comparison, the black continuous line shows the data for a sample without nanoparticles prepared with $\dot{T}=1 \, ^{\circ}$C/min.}
\label{fig:variousNP}
\end{figure}

For samples with NPs but prepared at lower heating rates, the concordance between the SANS and the visco-elasticity measurements suggests that the loss modulus is directly related to the total amount of NPs in the GBs, and thus that dissipation processes mainly involve the NPs confined in the grain boundaries. These processes are likely to be influenced by any interaction between the polymer and the NPs, such that the surface chemistry of the NPs should play a role. For the silica NPs used here, it is known that the copolymer F108 slightly adsorbs to the particle surface~\cite{Shar1999, Tamborini2012}. Note, in addition, that the fact that the structure peak originating from the NPs accumulated in the GBs is also visible for a SANS profile measured in silica-matched conditions (red squares in fig.~\ref{fig:SANS}), is also an indication of interactions between the F108 polymer and the NPs. We test the effect of the NPs-F108 interactions by performing a few additional visco-elasticity measurements using a different kind of nanoparticles, carboxylated polystyrene latexes, of diameter $36$ nm, comparable to that of the silica NPs. The effect of the polystyrene NPs on the sample structure, partial accumulation in the GBs and influence on the microstructure, is similar to that of the silica NPs~\cite{Ghofraniha2012}. Moreover, experiments on dilute suspensions of polymer and polystyrene NPs indicate an adsorption of the central block of the polymer to the surface of the NPs~\cite{Li1994} with an amount of adsorbed polymer per unit area of NPs comparable to the one evaluated for silica NPs ($1 \, \rm{mg/m}^2$)~\cite{Tamborini2012}. Despite these similarities, the impact of polystyrene NPs on the sample dissipation is strikingly different to that on silica NPs. This is illustrated in fig.~\ref{fig:variousNP}, where the frequency-dependent loss modulus is plotted for two heating rates $\dot{T}=1 \, ^{\circ}$C/min and $\dot{T}=0.01 \, ^{\circ}$C/min, for the two kinds of NPs.  The significant increase of the loss modulus at low $f$ observed for silica NPs-laden samples prepared at a small heating rate is not found for samples comprising polystyrene NPs. By contrast, we find that, for a sample with polystyrene NPs, $G"$ is comparable for samples prepared with a fast heating rate and with a slow heating rate. Hence at high rates, the behavior of samples comprising silica or polystyrene NPs are comparable, whereas they markedly differ at a low rate. This is consistent with the general picture that at low rate dissipation occurs mainly in the GBs (where NPs-polymer interactions are important), while at high rate NPs are more evenly dispersed in the sample and act as point defects that stabilize the sample. These preliminary, intriguing results deserve a deeper investigation that is out of the scope of the present work. They nevertheless illustrate nicely the crucial role of the interaction between the polymeric micelles and the NPs.

\section{Conclusion}

We have investigated how impurity-like nanoparticles added in a colloidal crystal impact its linear visco-elasticity, because of their partial segregation in the grain-boundaries that separate crystalline grains with different orientations. Two independent experimental quantities, the volume fraction of NPs and the heating rate that controls the crystallization, have been varied, both allowing the microstructure of the polycrystal to be tuned. We have found that the elasticity is independent of the presence of nanoparticles and the average size of the crystallites. In sharp contrast, the dissipation strongly varies with the presence of NPs and with the sample microstructure. By combining rheology measurements and structural measurements, we conclude that the viscous dissipation is governed by the cumulated amount of nanoparticles located in the grain boundaries. We are not aware of previous comparable results for colloidal crystals. A comparison can nevertheless be drawn with other soft materials.
How the properties of the interface impact the loss modulus has for instance been investigated in foams~\cite{Marze2009}. Interestingly, in particles-laden foams the loss modulus $G"$ has been observed to increase with the amount of particles~\cite{CohenAddad2007}, indicating, as in our case, that dissipation processes are correlated to the presence of nanoparticles in the interface. However, in particles-laden foams the storage modulus increases as well with the amount of particles, because the elasticity of foams is also directly related to the interfacial properties. By contrast, in our polycrystals elasticity and dissipation are governed by different structural features, the packing of the micelles for the former, and the  amount of NPs in the grain boundaries for the latter. In this respect, colloidal polycrystals doped with impurities share more analogies with their atomic and molecular counterparts. The fact that impurity segregation in the grain-boundaries modifies the mechanical properties of the polycrystal is indeed reminiscent of the embrittlement of metal by solute segregation in the grain boundaries that is eventually responsible for intergranular fracture leading to material failure~\cite{Duscher2004, Schweinfest2004, Chen2010}.




\section{Acknowledgement}

This work has been supported by ANR
under Contract No. ANR-09-BLAN-0198 (COMET), and by Laboratoire L\'eon Brillouin (LLB). We thank F. Cousin (LLB) for help with the small-angle
neutron scattering measurements. Discussion with N. Ghofraniha, F. Caton and D. Roux are gratefully acknowledged.



\end{document}